\documentclass{elsart}
\usepackage{epsfig}
\usepackage{rotating}
\begin{document}
\runauthor{Marco Battaglia}
\begin{frontmatter}
\title{Cluster Imaging\\ with a Direct Detection CMOS Pixel Sensor\\ 
in Transmission Electron Microscopy}
\author[UCB,LBNL]{Marco Battaglia\corauthref{cor}},
\corauth[cor]{Corresponding author, Address: Lawrence Berkeley National Laboratory, 
Berkeley, CA 94720, USA, Telephone: +1 510 486 7029.} 
\ead{MBattaglia@lbl.gov}
\author[LBNL]{Devis Contarato},
\author[LBNL]{Peter Denes},
\author[LBNL,Padova]{Piero Giubilato}
\address[UCB]{Department of Physics, University of California at 
Berkeley, CA 94720, USA}
\address[LBNL]{Lawrence Berkeley National Laboratory, 
Berkeley, CA 94720, USA}
\address[Padova]{Dipartimento di Fisica, Universit\`a degli Studi, 
Padova, I-35131, Italy}
\begin{abstract}
A cluster imaging technique for Transmission Electron Microscopy with
a direct detection CMOS pixel sensor is presented. Charge centre-of-gravity 
reconstruction for individual electron clusters improves the spatial 
resolution and thus the point spread function. Data collected with a CMOS 
sensor with 9.5$\times$9.5~$\mu$m$^2$ pixels show an improvement of a factor 
of two in point spread function to 2.7~$\mu$m at 300~keV and of a factor of 
three in the image contrast, compared to traditional bright field 
illumination.

\end{abstract}
\end{frontmatter}

\typeout{SET RUN AUTHOR to \@runauthor}


\section{Introduction}

The use of radiation-tolerant CMOS active pixel sensors (APS)~\cite{fossum, cmos} 
for direct detection in Transmission Electron Microscopy (TEM), opens new opportunities 
for fast imaging with high sensitivity~\cite{emicro, deptuch, Denes2007,fan,Battaglia:2008yt}. 
One of the key figures of merit for a TEM imaging sensor is the point spread function (PSF), 
which affects both the imaging resolving capabilities of the sensor and the absolute image
contrast~\cite{thust}.
The PSF of a pixellated APS used in direct detection depends on several parameters of which 
the most important are the pixel size, the electron multiple scattering in the sensor and 
the charge carrier diffusion in the active layer. This letter discusses the PSF achieved 
using an APS in two different regimes.

In traditional bright field illumination, the electron flux is such that each pixel 
is illuminated by one or more electrons per acquisition frame. 
In this regime the point spread function has a contribution from the lateral 
charge spread due to charge carrier diffusion in the active volume. Since the 
epitaxial layer of a CMOS APS is nearly field-free, charge carriers reach the 
collection diode through thermal diffusion, with collection times of 
${\cal{O}{\mathrm{(150~ns)}}}$. The typical cluster size for a CMOS APS with 
10~$\mu$m pixels is 4-5~pixels for 300~keV electrons and about 45~\% of the charge 
is collected on the central pixel of the cluster. With bright field illumination, 
at high rate, the signal recorded on each individual pixel is the superposition 
of the charge directly deposited by a particle below the pixel area with that 
collected from nearby pixels through diffusion, multiple scattering and 
backscattering from the bulk Si.
If the electron rate is kept low enough so that individual electron clusters can 
be reconstructed, a new regime of operation becomes available. The electron impact 
position is reconstructed by calculating the centre-of-gravity of the observed pulse 
heights on the pixels in the cluster.  For pixel detectors with 
$\cal{O}{\mathrm{(10~\mu m)}}$ pixel pitch this technique makes it possible to 
obtain an $\cal{O}{\mathrm{(1~\mu m)}}$ point resolution in tracking applications 
for accelerator particle physics. The same technique can now be adopted for imaging, 
provided electron fluxes are low enough so that the detector occupancy is $\le$~0.05, 
and individual clusters can easily be resolved. 
Under these conditions the PSF is expected to depend only on the 
detector pixel size and cluster S/N (determining the single point resolution) and on 
multiple scattering. The image is reconstructed by combining a large number of frames, 
each giving 2-5~$\mu$m accuracy for 0.01-0.05~\% of the field of view, provided the frame 
rate is much faster than the dynamics being observed on the sample. In this paper we 
refer to this technique as ``cluster imaging''.

\section{Cluster Imaging Tests}

The principle of cluster imaging for TEM has been tested using the TEAM1k CMOS monolithic
pixel sensor, developed as part of the the multi-institutional TEAM (Transmission Electron 
Aberration-corrected Microscope) project~\cite{team}. The TEAM1k sensor and its performance 
will be discussed in details in a forthcoming paper on this Journal~\cite{team1k}. 
The chip is produced in the AMS 0.35~$\mu$m CMOS-OPTO process and has a 1,024$\times$1,024 
pixel imaging area  with pixels arrayed on a 9.5~$\mu$m pitch. The detector employed in this 
test has been back-thinned to 50~$\mu$m and mounted on a carrier circuit board which is cut out 
below its active area to minimise back-scattering effects. A 75~$\mu$m-diameter Au wire is 
mounted along the pixel rows on top of the sensor at a distance of $\sim$2~mm above 
its surface. Tests have been conducted with the sensor installed at the bottom of a FEI 
Titan microscope column at the National Center for Electron Microscopy (NCEM).

The energy deposited by electrons in the sensor active layer and the lateral 
charge spread are simulated with the {\tt Geant-4} program~\cite{Agostinelli:2002hh}
using the low-energy electromagnetic physics models~\cite{Chauvie:2001fh}.
The CMOS pixel sensor is modelled according to the detailed geometric structure 
of oxide, metal interconnect and silicon layers.
Electrons are incident perpendicular to the detector plane and tracked through it. 
Charge collection in the sensor is simulated with {\tt PixelSim}, a dedicated 
digitisation module~\cite{Battaglia:2007eu}. The simulation has the diffusion 
parameter $\sigma_{{\mathrm{diff}}}$, used to determine the width of the charge 
carrier cloud, free. Its value is extracted from data by a $\chi^2$ fit to the pixel 
multiplicity in the clusters of 300~keV electrons, where multiple scattering 
is lower. We find $\sigma_{{\mathrm{diff}}}$ = (14.5$^{+2.0}_{-1.0}$)~$\mu$m, 
which is compatible with the value obtained for 1.5~GeV $e^-$s in a prototype 
sensor with 20$\times$20~$\mu$m$^2$ pixels produced in the same CMOS 
process~\cite{Battaglia:2009aa}. 
The response to single electrons is characterised in terms of the 
cluster size. We operate with a flux of $\simeq$~50~$e^-$~mm$^{-2}$~frame$^{-1}$ 
which allows us to resolve individual electrons. Electron hits are 
reconstructed as clusters of pixels. A clustering algorithm with two 
thresholds is used~\cite{Battaglia:2008yt}. First the detector is scanned 
for ``seed'' pixels with pulse height values over S/N threshold set to 3.5. 
Seeds are sorted according to their pulse heights and the surrounding 
neighbouring pixels are added to the cluster, if their S/N exceeds 2.5.
\begin{figure}
\begin{center}
\begin{tabular}{c c}
\epsfig{file=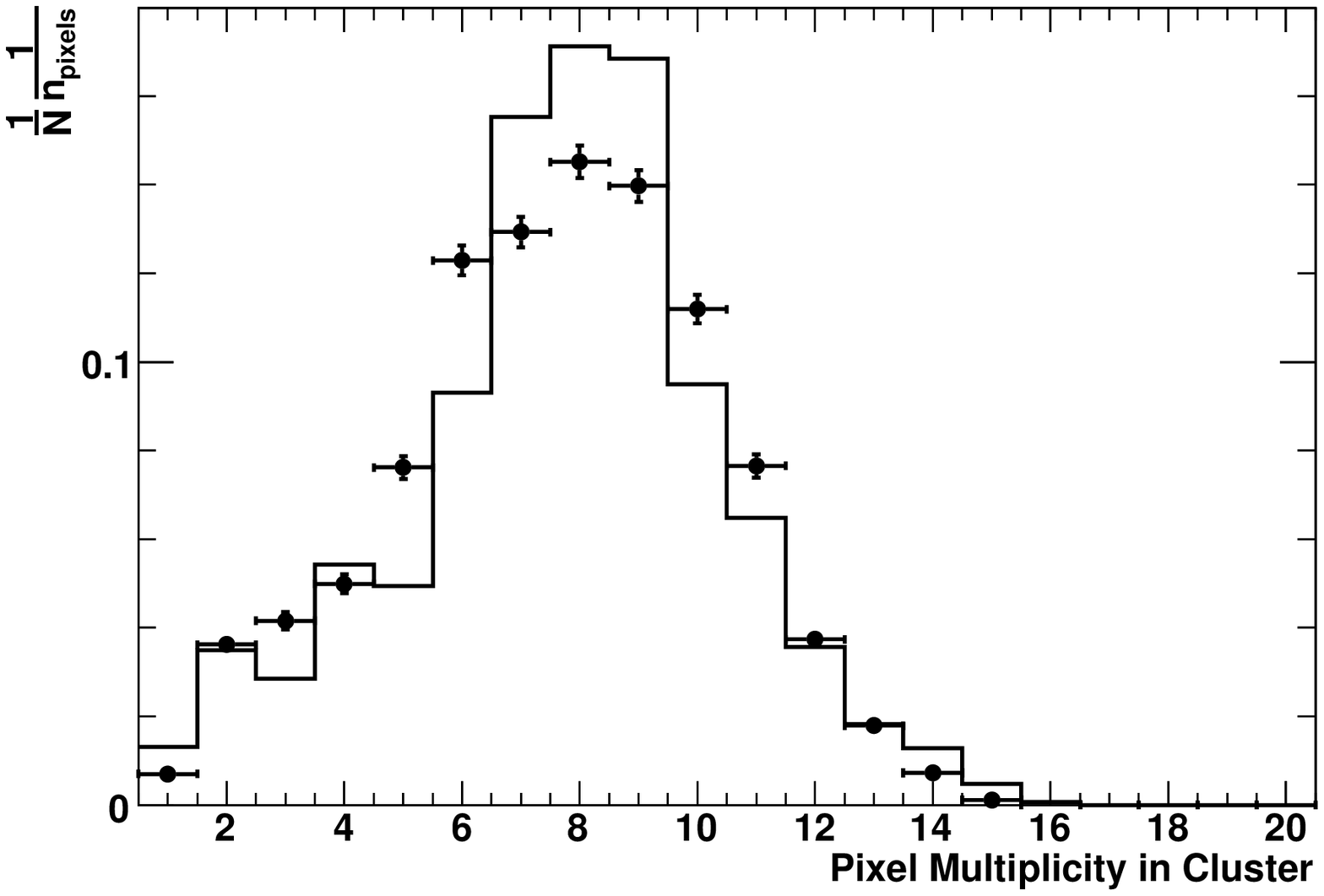,width=6.75cm,clip=} &
\epsfig{file=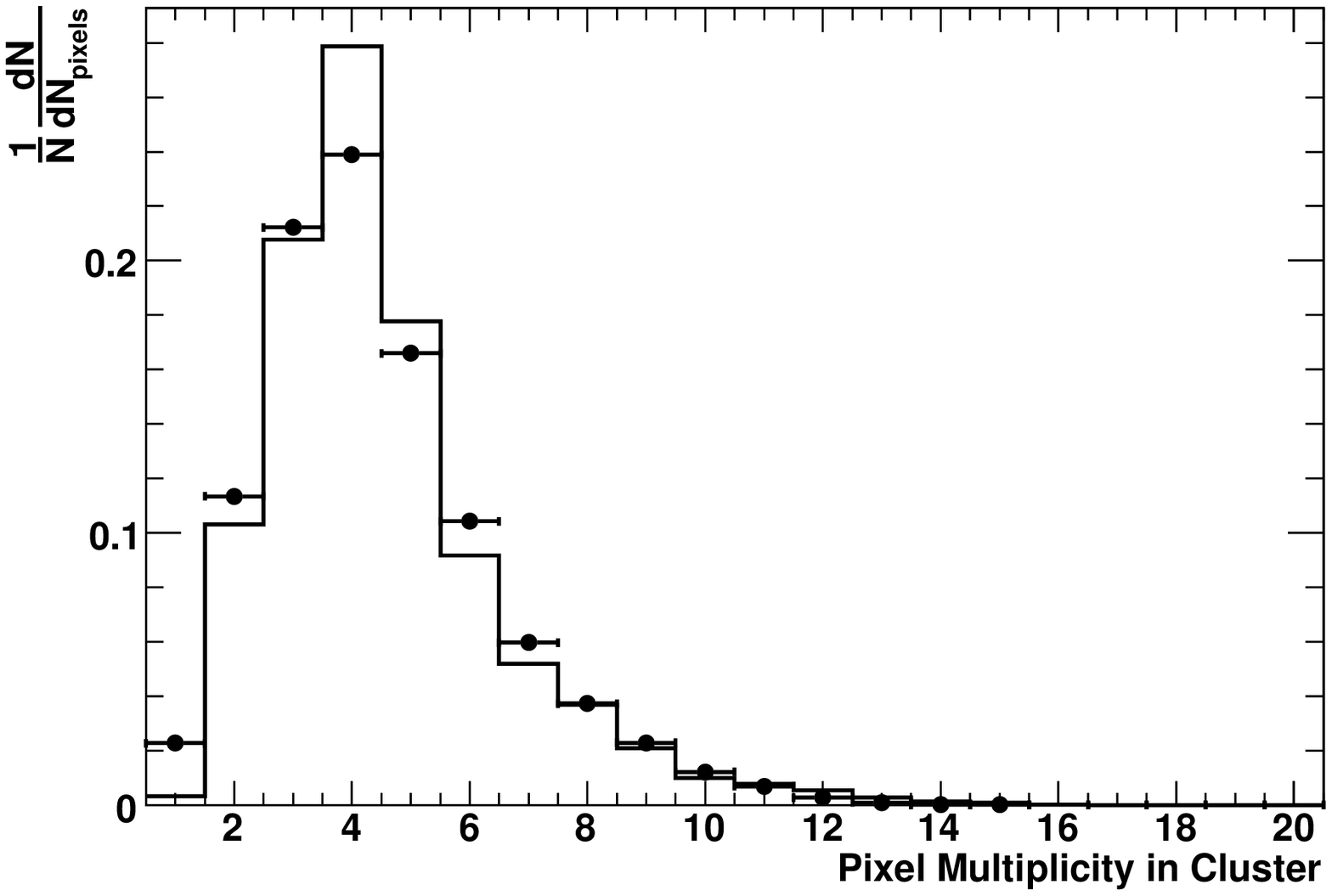,width=6.75cm,clip=} \\
\end{tabular}
\end{center}
\caption{Number of pixels in single electron clusters for 80~keV 
(left panel) and 300~keV (right panel) electrons. The points 
with error bars show the data and the line the result of the 
{\tt Geant-4}+{\tt PixelSim} simulation.}
\label{fig:npixels}
\end{figure}
The pixel multiplicity at different electron energies is shown 
in Figure~\ref{fig:npixels}. The agreement of the tuned {\tt PixelSim} 
simulation with data is good and we observe an increase of multiplicity
at lower energies, as expected from the increased multiple scattering 
and energy deposition. The electron point of impact is determined from the 
position of the pixels in the cluster weighted by their pulse height.
The average cluster S/N in data is 17 at 300~keV and 42 at 80~keV, due 
to the increase of the electron energy loss. The electron position of 
impact is computed as the centre-of-gravity of the charge recorded in the cluster.
The {\tt Geant-4} + {\tt PixelSim} simulation, which correctly reproduces the 
cluster signal pulse height and pixel multiplicity, predicts a point 
resolution for this reconstruction method of (6.75$\pm$0.15)~$\mu$m  and 
(2.80$\pm$0.04)~$\mu$m for 80~keV and 300~keV electrons, respectively. 
As the energy decreases, multiple scattering in the inactive layers above the 
epitaxial layer increases, degrading the PSF.  Better position resolution is 
obtained in particle physics applications, as the particles are more energetic.

\begin{figure}
\begin{center}
\epsfig{file=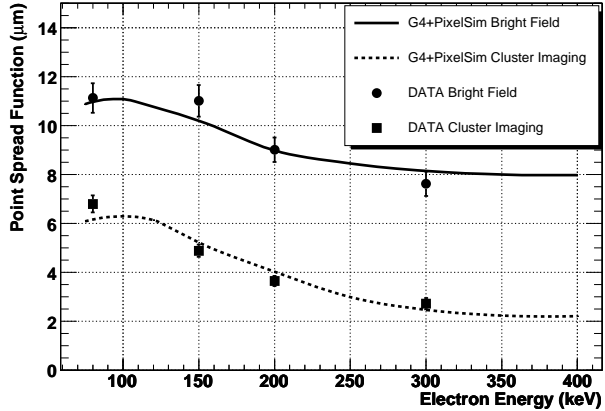,width=9.0cm,clip=}
\end{center}
\caption{Fitted line spread function as a function of beam energy. 
 Cluster imaging results (squares with error bars) are compared to 
 bright field pixel illumination (circles with error bars) and 
 simulation predictions (lines).}  
\label{fig:plotPSF}
\end{figure}

We compare the cluster imaging technique to bright field illumination, using 
the same detector, as a function of the electron energy. 
The PSF is measured on data along one coordinate from the image projected onto the 
sensor by the thin Au wire as discussed in~\cite{Battaglia:2008yt}. Since the Au wire 
has well-defined edges, the profile of the deposited energy in the pixels, measured 
across the wire in bright field illumination allows us to study the lateral 
charge spread due to scattering and diffusion along the projected image of the edges. 
The electron flux is approximately 5$\times$10$^3$~e$^-$~mm$^{-2}$~frame$^{-1}$.
The line spread function (LSF) is extracted by parameterising the measured pulse height 
on pixel rows across the wire image with a box function smeared by a LSF Gaussian term. 
The average pulse height level for pixels away from the wire region and the contrast 
ratio to the response below the wire are fixed to those observed in data and the LSF value 
is obtained by a simple 1-parameter $\chi^2$ fit with the Gaussian width as free parameter. 
At 300~keV, we measure a LSF value of (7.6$\pm$0.5)~$\mu$m which becomes 
(11.1$\pm$0.6)~$\mu$m at 80~keV, where the uncertainty include statistics and 
systematics from the result stability across the columns of a sector.
\begin{figure}
\begin{center}
\begin{tabular}{c c}
\epsfig{file=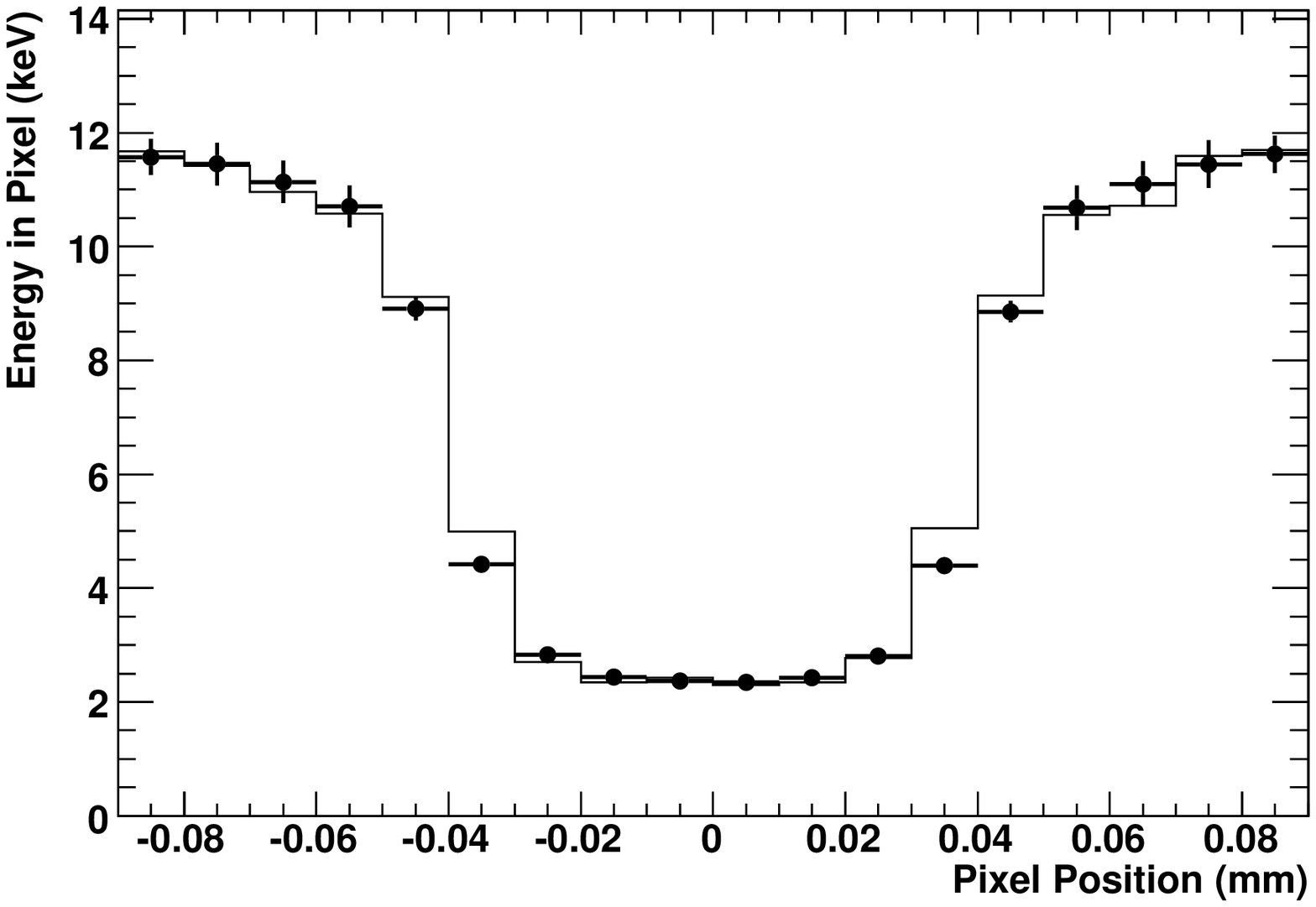,width=6.75cm,clip=} &
\epsfig{file=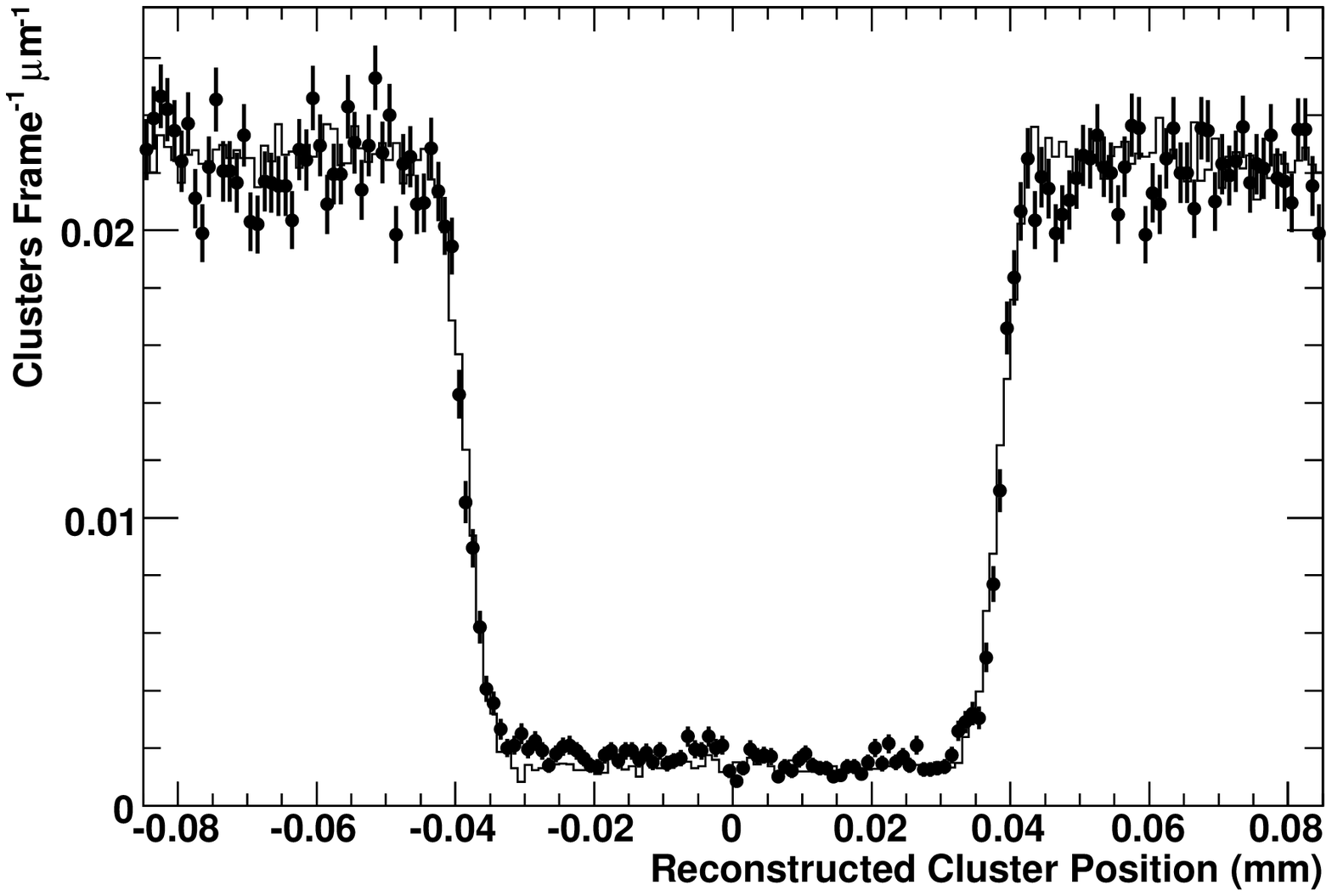,width=6.75cm,clip=} \\
\end{tabular}
\end{center}
\caption{Au wire image on the TEAM1k detector for 300~keV electrons:
(left panel) energy recorded in pixel vs.\ pixel position across 
the wire with bright field illumination, (right panel) number of 
clusters reconstructed per frame and micron bin vs.\  position across 
the wire for cluster imaging. The points with error bars represent
the data and the continuous lines the result of the fit. The fitted 
values of the line spread function are (7.6$\pm$0.5)~$\mu$m and 
(2.72$\pm$0.25)~$\mu$m, respectively.}
\label{fig:plotWire}
\end{figure}
For cluster imaging, the electron flux is reduced to 
$\simeq$~50~e$^-$~mm$^{-2}$~frame$^{-1}$ and 20000 subsequent frames are acquired 
at each energy. Figure~\ref{fig:plotWire} shows the image of the wire obtained in the 
two regimes. With cluster imaging, due to the continuous spatial sampling, the information 
is no longer discretised by the pixel pitch as it is with bright field illumination. 
Charge diffusion, which limits the PSF in bright field illumination, now rather 
improves the cluster resolution by spreading the charge on more pixels and the 
relevant parameter for PSF becomes the ratio of pixel pitch to signal-to-background, 
which determines the cluster spatial resolution. We also observe that the 
image contrast improves by a factor of three for cluster imaging compared to 
bright field illumination, as it is apparent in Figure~\ref{fig:plotWire}.   
The LSF is extracted by a $\chi^2$ fit, as discussed above. We measure 
(2.72$\pm$0.25)~$\mu$m at 300~keV, (3.65$\pm$0.20)~$\mu$m at 200~keV and 
(6.80$\pm$0.35)~$\mu$m at 80~keV. These results show that cluster imaging 
improves the point spread function by a factor of two compared to traditional 
bright field illumination, for the TEAM1k detector parameters.
We perform the study on simulation describing the detector and wire geometry, 
for both operation regimes and compare the results to the measured LSF values. 
Results are summarised in Figure~\ref{fig:plotPSF}. Simulation reproduces well 
the data both in absolute terms and in the observed scaling with the electron 
energy. We also note that the cluster imaging LSF closely follows the simulation 
predictions for single point resolution. 

\section{Conclusions}

The TEAM CMOS APS has demonstrated a PSF for direct detection under bright field 
illumination comparable to its pixel size. Image reconstruction by clustering at 
the impact point of individual electrons at low flux allows to reduce the PSF down 
to (2.72$\pm$0.25)~$\mu$m and improves the image contrast. In order to practically 
exploit this method in TEM it is necessary to ensure that the sensor can be read 
out at a rate of several hundred frames~s$^{-1}$ and cluster reconstruction can be 
performed in real time.

\section*{Acknowledgements}

\vspace*{-0.1cm}

This work was supported by the Director, Office of Science, 
of the U.S. Department of Energy under Contract No.DE-AC02-05CH11231.
The TEAM Project is supported by the Department of Energy, Office of 
Science, Basic Energy Sciences.

\vspace*{-0.1cm}

 \nolinenumbers

\end{document}